\documentclass[12pt]{iopart}
\usepackage{graphicx} 
\begin{document}
\title{Optimum Monte Carlo Simulations: Some Exact Results}
\author{J. Talbot$^1$, G. Tarjus$^2$ and P. Viot$^2$}
\address{$^1$Department   of  Chemistry  and  Biochemistry,
 Duquesne University, Pittsburgh, PA 15282-1530\\ 
$^2$Laboratoire  de  Physique Th\'eorique  des Liquides,  Universit\'e
Pierre et Marie Curie, 4, place Jussieu, 75252 Paris Cedex, 05 France}

\begin{abstract}
We obtain exact  results for  the  acceptance ratio  and mean  squared
displacement  in   Monte Carlo  simulations   of the  simple  harmonic
oscillator  in $D$  dimensions.  When  the  trial displacement is made
uniformly    in  the radius,  we  demonstrate    that the results  are
independent  of the dimensionality of  the  space.  We  also study the
dynamics  of  the process via  a  spectral  analysis and  we obtain an
accurate description for the relaxation time.
\end{abstract}
\section{Introduction}

Since the original Metropolis  algorithm  appeared five  decades  ago,
countless studies   have employed the     technique  to evaluate   the
thermodynamic properties of    model systems\cite{AT87,B97,FS02}.  The
essence of the method is to generate a sequence of configurations that
represent    a  given  thermodynamic  ensemble,   often  the canonical
ensemble.  Properties of interest are  then obtained as averages  over
the   configurations.  At  each step    of   the simulation a    trial
configuration  is  obtained from the current   one  by making a random
displacement in the configuration space.  This might correspond to, for
example, displacing   a   randomly  selected  particle.     The  trial
configuration is either accepted or  rejected with a probability given
by the appropriate   Boltzmann factor for   the ensemble. In case   of
rejection, the current configuration is retained for use in evaluating
the properties of interest.

Since many applications of  MC are  computationally intensive, a  much
addressed issue   has  been the  optimization  of  the simulation with
respect to one or more   control parameters so that the  configuration
space  is   sampled in the  most efficient   way.  Bouzida,  Kumar and
Swendsen  (BKS)\cite{BKS92,S02}, as   part    of a program   aimed  at
improving the efficiency of MC  simulations of biomolecules, performed
numerical studies of the  simple harmonic  oscillator (SHO) where  the
convergence  of    the   simulation      depends  on    the    maximum
displacement. There is no unique measure of efficiency, but two simple
choices are the mean  squared and mean  absolute displacements. As the
maximum displacement tends  to zero or infinity  it is clear  that the
average value of both of these quantities tends  to zero since, in the
first case   the particle  does not  move,  while in  the   second all
attempted moves are rejected.  Thus both quantities have a maximum for
some intermediate value of the maximum displacement.

It is also useful to consider the dynamical process associated with an
MC simulation,  even  though it  does  not  correspond  to the  actual
dynamics of  the system. One can,  for example, calculate various time
correlation functions that  can  be    used to develop     alternative
efficiency criteria\cite{K88,MT94}.

BKS\cite{BKS92} performed numerical studies of the SHO in one, two and
three  dimensions and examined  the  acceptance ratio $P_{acc}$ (i.e.,
the fraction  of accepted trial   configurations), mean squared,  $<(\Delta
x)^2)>$, and mean absolute, $<|\Delta x|>$,  displacements as a function of
the  maximum displacement, $\delta$.  They found  that the acceptance ratio
decreases approximately exponentially for small to intermediate values
of $\delta$  and then inversely for larger  values.  In  one dimension they
found  that the   maxima  in $<(\Delta  x)^2>$    and $<|\Delta x|>$   occur  at
$P_{acc}=0.42$ and $P_{acc}=0.56$, respectively.  In higher dimensions
the results depend on how the jump is made.  BKS considered two cases:
in one the jumps are  performed uniformly to  any point in a spherical
volume of  radius $\delta$ centered on  the current position, a choice that
favors  larger radial displacements for dimensions  $D>1$. In a second
method,  the jumps were sampled  uniformly in the radius (and randomly
in  the  orientation) so  that  all radial  displacements are  equally
probable. In the former case BKS observed that, for a given $\delta$, $P_{acc}$
decreases as a function of $D$, while  for uniform radius sampling the
numerical results suggested that $P_{acc}$ is independent of $D$.

In addition to these     static  properties, BKS also  examined    the
correlation time, $\tau$, of the energy-energy correlation function. They
observed  a  minimum  correlation  time for  an   acceptance  ratio of
approximately $50\%$.

Here we present an analytical study of the SHO in arbitrary dimension.
We obtain exact expressions   for the acceptance   ratio and the  mean
squared and mean  absolute displacements as  functions  of the maximum
displacement $\delta$.    We show  that when   the trial jump   is selected
uniformly in the radius, the results are independent of the dimension.
We also present an analysis of the dynamics of the process.

\section{ONE DIMENSION}
We first   investigate the  case of   a SHO  in  one  dimension, whose
potential energy is given by  $V(x)=kx^2/2$ where $x$ is the  position
and $k$ is the stiffness constant.

In a standard Metropolis Monte Carlo Simulation one makes a trial move
with a uniform random displacement selected between $-\delta$ and $\delta$.
The dynamical process generated by the successive trial moves of the
Monte Carlo simulation can be written
\begin{eqnarray}\label{eq:1}
\fl\frac{dP(x,t)}{dt}&=-\frac{1}{2\delta}\int_{-\delta}^{\delta}dhW(x\to x+h)P(x,t)+
 \frac{1}{2\delta}\int_{-\delta}^{\delta}dhW(x+h\to x)P(x+h,t)
\end{eqnarray}
where $W(x\to x+h)$  denotes the transition rate  from the state $x$  to
the state $x+h$ and $P(x,t)$ is the probability to find the oscillator
at   position $x$  at time $t$.     To ensure the convergence  towards
equilibrium, a  sufficient condition  is  given by  detailed  balance,
which is expressed as
\begin{equation}\label{eq:2}
\frac{W(x\to x+h)}{W(x+h\to x)}=\frac{P_{eq}(x+h)}{P_{eq}(x)}
\end{equation}
where
\begin{equation}\label{eq:3}
P_{eq}(x)=c \exp(-\beta V(x))
\end{equation}
and $c=\sqrt{\beta k/2\pi}$ is a normalization constant ensuring that
$\int_{-\infty}^{\infty} dx
P_{eq}(x)=1$.  One solution of Eq.~(\ref{eq:2}) is the Metropolis
rule,
\begin{equation}\label{eq:4}
W(x\to x+h)=\min(1,\exp(-\beta(V(x+h)-V(x))).
\end{equation}

\begin{figure}
\begin{center}
\resizebox{10cm}{!}{\includegraphics{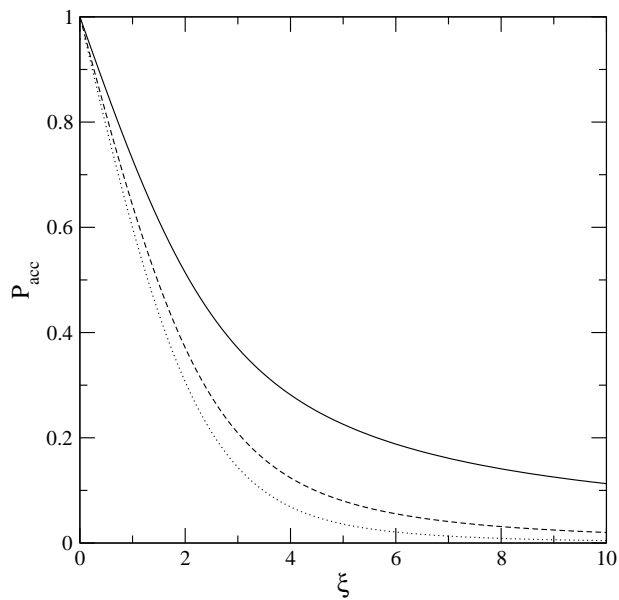}}

\caption{Acceptance ratio $P_{acc}$ versus $\xi=\sqrt {\frac{\beta k}{2}}\delta$ 
for a harmonic oscillator with uniform volume sampling in $D=1,2$, and
$3$ dimensions (full, dashed,  and  dotted lines respectively). For  a
uniform radius sampling, all  curves coincide with the one-dimensional
result (full line).}\label{fig:1}
\end{center}
\end{figure}

Most of  the properties  of the $1D$  SHO can  be obtained analytically.
For instance,  the acceptance ratio, which is  the number  of accepted
trials over the total number of trials can be expressed as
\begin{equation}\label{eq:5}
P_{acc}(\delta)=c\int_{-\infty}^{+\infty}  dxe^{-\beta  V(x)}   \frac{1}{2\delta}\int_{-\delta}^{\delta}dhW(x\to
x+h).
\end{equation}
In   this  equation $\exp(-\beta  V(x))dx$  is  the   probability that the
oscillator is  between $x$ and $x+dx$,  $dh/2\delta$ is  the probability of
selecting a random displacement between $h$ and $h+dh$ and $W(x\to x+h)$
is the  probability of  accepting  the  trial displacement  (given  by
Eq.~(\ref{eq:4})).  Integration over  the  allowed values of  $x$ and
$h$   then    gives  the   average     acceptance probability.   Since
displacements to  the left and right  are symmetric,  we need consider
only one direction.  For displacements to  the right, Eq.~(\ref{eq:5}),
can be written as
\begin{eqnarray}\label{eq:6}
\frac{d(\delta   P_{acc}(\delta ))}{d\delta  }&=c\int_{-\infty }^{+\infty  }dx e^{-\beta V(x)} W(x\to x+\delta).
\end{eqnarray}
For $x<-\delta/2$, $W(x\to x+\delta)=1$ and for $x>-\delta/2$, $W(x\to x+\delta
)=e^{-\beta k((x+\delta)^2-x^2)/2}$. One thus obtains
\begin{eqnarray}\label{eq:7}
\frac{d(\xi    P_{acc}(\xi  ))}{d\xi   }&=1-erf\left(\frac{\xi}{2} \right)
\end{eqnarray}
where $erf(x)$ is the error  function and $\xi=\sqrt{\frac{\beta k}{2}}\delta $. Using the
initial  condition, i.e., $P_{acc}(0)=1$,   the   solution of the
differential equation~(\ref{eq:7}) is
\begin{equation}
P_{acc}(\xi )=1-erf\left(\frac{\xi}{2} \right)+\frac{2}{\sqrt{\pi}\xi }\left(1-e^{-\xi^2/4}\right).
\end{equation}
The function is plotted in Fig.~\ref{fig:1}.

\begin{figure}
\begin{center}

\resizebox{9cm}{!}{\includegraphics{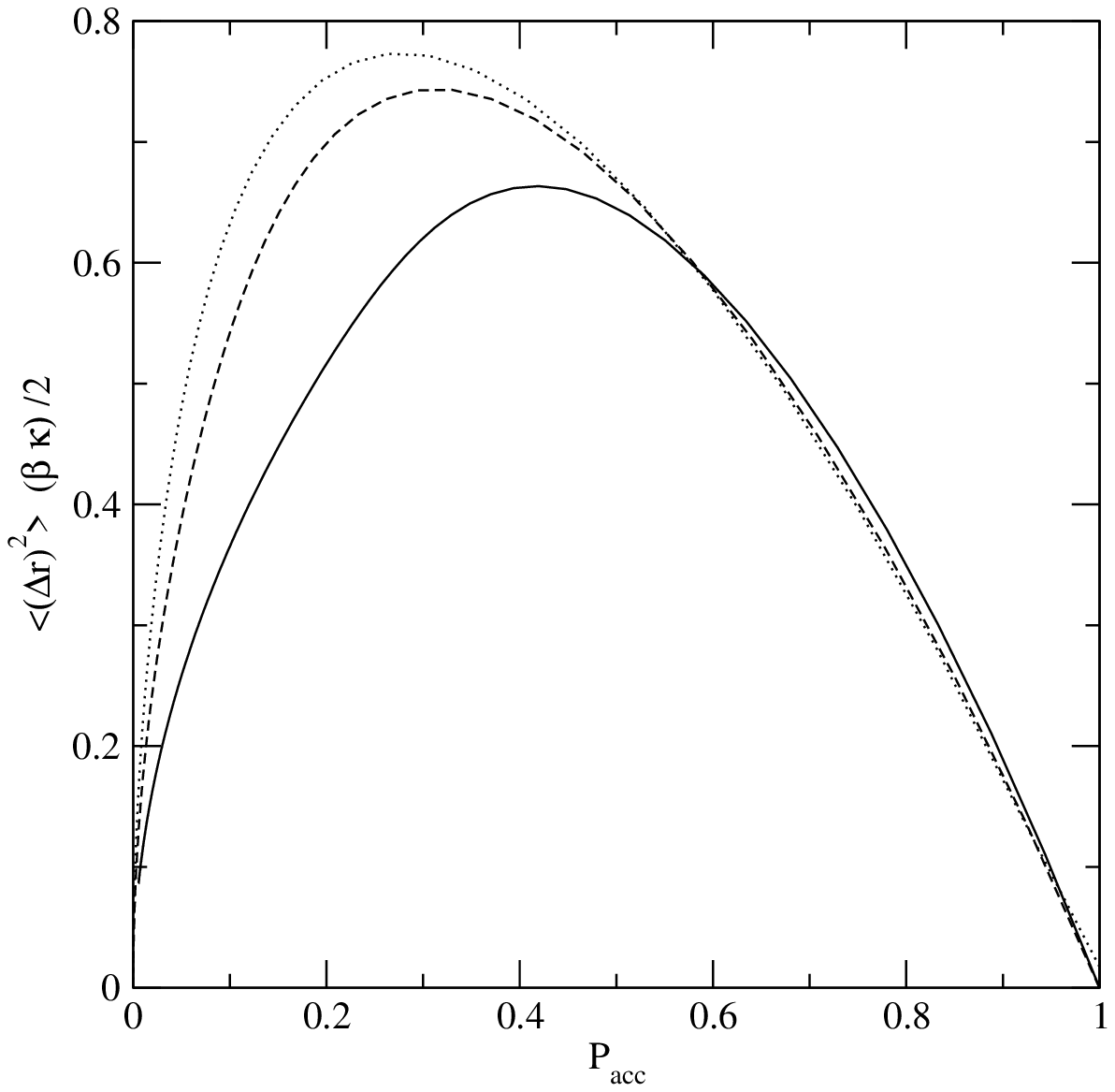}}
\caption{Mean squared displacement $<(\Delta r)^2>\frac{\beta k}{2}$
versus  the acceptance ratio  $P_{acc}$ for uniform volume sampling in
one, two and three  dimensions (full curve,  dashed, and dotted lines,
respectively). For a uniform radius sampling  all curves coincide with
the one-dimensional result (full line). }\label{fig:2}
\end{center}
\end{figure}

\begin{figure}
\begin{center}
\resizebox{9cm}{!}{\includegraphics{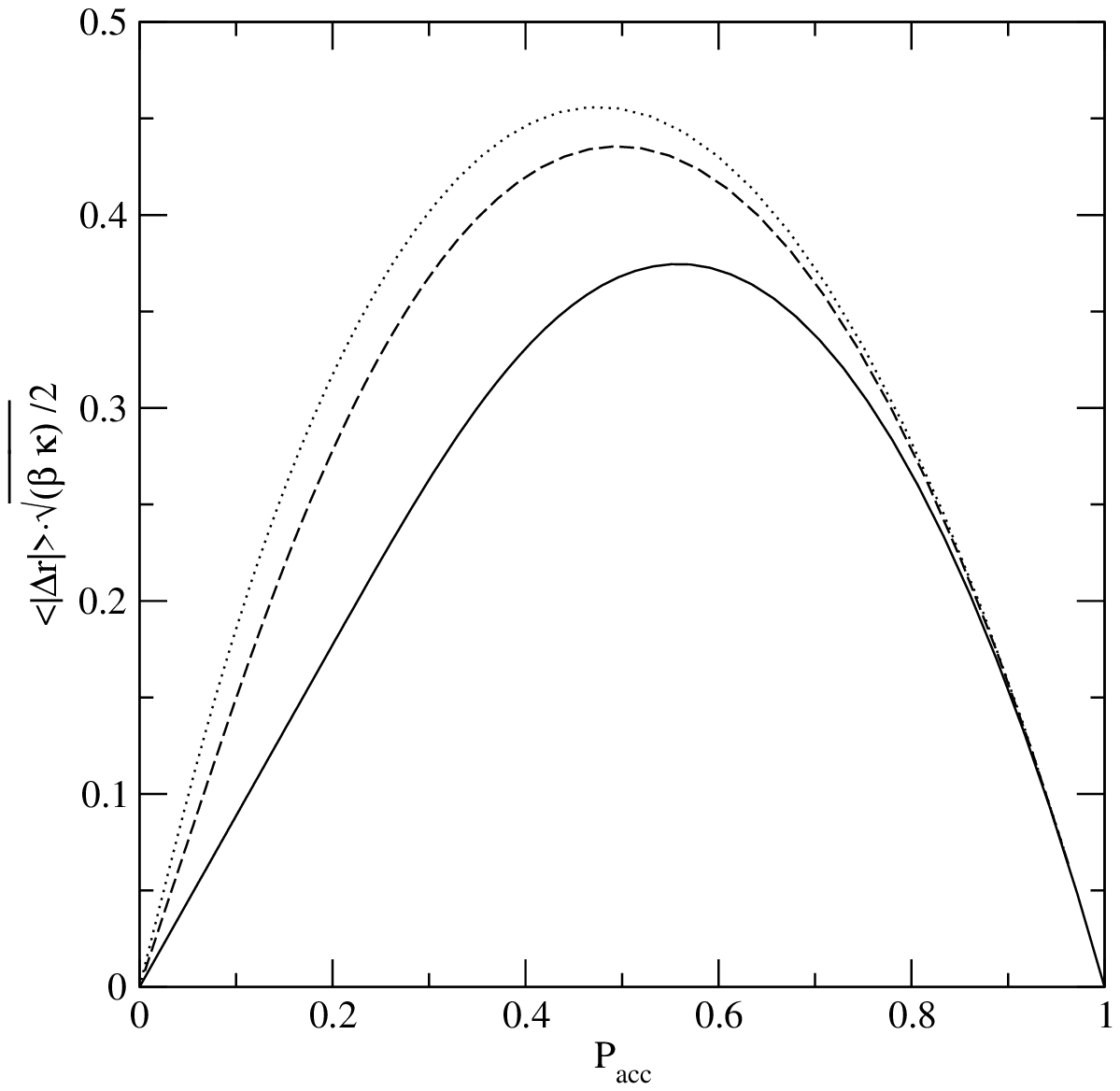}}

\caption{Mean absolute  displacement $<|\Delta r|>\sqrt{\frac{\beta k}{2}}$
versus the acceptance ratio $P_{acc}$ for a uniform volume sampling in
one, two and  three dimensions (full curve,  dashed, and dotted lines,
respectively). For a uniform radius sampling  all curves coincide with
the one-dimensional result (full line).}\label{fig:3}
\end{center}
\end{figure}
The mean  squared displacement,  $<(\Delta  x)^2>$,  and the  mean absolute
displacement, $<|\Delta x|>$, defined as
\begin{eqnarray}
<(\Delta x)^2>&=c\int_{-\infty}^{+\infty}  dx\exp(-\beta V(x))
\frac{1}{2\delta}\int_{-\delta}^{\delta}dhh^2W(x\to x+h),\\
<|\Delta x|>&=c\int_{-\infty}^{+\infty}  dx\exp(-\beta V(x))
\frac{1}{2\delta}\int_{-\delta}^{\delta}dh|h|W(x\to x+h).
\end{eqnarray}
can be quite simply obtained from the generating function
\begin{equation}\label{eq:30}
Z(\lambda)=c\int_{-\infty}^{+\infty}  dx\exp(-\beta V(x))
\frac{1}{2\delta}\int_{-\delta}^{\delta}dh\exp(-\lambda|h|)W(x\to x+h),
\end{equation}
by the derivatives with respect to $\lambda$ and set $\lambda=0$, i.e., $<|\Delta
x|>=-(\partial Z(\lambda)/\partial\lambda)_{\lambda=0}$, $<(\Delta x)^2>=(\partial^2Z(\lambda)/\partial\lambda^2)_{\lambda=0}$ (and, of
course, $P_{acc}=Z(\lambda=0)$).

By multiplying    both sides of  Eq.~(\ref{eq:30})   by  $\xi$  and next
differentiating with respect to $\xi$, one obtains

\begin{equation}\label{eq:34}
\frac{\partial(\xi  Z(\lambda, \xi))}{\partial\xi}=\exp\left(-\sqrt{\frac{2}{\beta k}}\lambda \xi\right)\left(1-erf\left(\frac{\xi}{2}\right)\right) 
\end{equation}
which, after defining $\tilde{\lambda}=\sqrt{\frac{2}{\beta k}}\lambda$, leads to
\begin{equation}\label{eq:8}
 \fl          Z(\lambda, \xi)=\frac{1}{\tilde{\lambda}\xi        }\left[1-     \exp(-{
 \tilde{\lambda}\xi})\left(1-                             erf\left(\frac{\xi}{2}
 \right)\right)-\exp(\tilde{\lambda}^2)\left(
 erf\left(\frac{\xi}{2}+\tilde{\lambda}\right)-erf(\tilde{\lambda})\right)\right]
\end{equation}

By taking the derivatives of the above formula with respect to $\lambda$ and
evaluating the resulting expressions at $\lambda=0$, it is easy to show that
the mean squared and mean absolute displacements are given by

\begin{equation}\label{eq:33}
\fl \sqrt{\frac{\beta k}{2}} <|\Delta x|>=\left[\frac{\xi}{2}\left(1- erf\left(\frac{\xi}{2} \right)\right)
-\frac{1}{\sqrt{\pi} }\exp\left(-\frac{\xi^2}{4}\right)+\frac{erf\left(\frac{\xi}{2}
\right)}{\xi}\right],
\end{equation}
\begin{equation}\label{eq:32}
\fl\frac{\beta k}{2} <(\Delta    x)^2>=\frac{1}{3}\left[\xi^2\left(1- erf\left(\frac{\xi}{2}  \right)\right)
+\frac{8}{\sqrt{\pi}\xi
}\left(1-\left(1+\frac{\xi^2}{4}\right)\exp\left(-\frac{\xi^2}{4}
\right)\right)\right].
\end{equation}

The maximum in the mean  squared and mean absolute displacements occur
for $\xi=2.61648$ and $\xi=1.76332$, which corresponds to acceptance ratio
values of  $P_{acc}=0.41767$ and  $P_{acc}=0.558239$ in agreement with
the  numerical results  of BKS\cite{BKS92}: see  Figs.~\ref{fig:2} and
~\ref{fig:3}.

\section{D dimensions}
We  show here that the acceptance  ratio, mean squared displacement and
other quantities   of    interest can be   obtained   exactly  in  any
dimension. Note that the term ``volume'' should be interpreted as the
hypervolume in $D$ dimensions, e.g., area in 2D and volume in 3D. 
For  simplicity, we   derive exact   expressions in  odd
dimensions, but similar results can be obtained in even
dimensions. 
\subsection{Uniform volume sampling}
In $D$ dimensions, the acceptance ratio is expressed as
\begin{equation}\label{eq:9}
\fl P_{acc}(\delta)=\frac{c_D}{\delta^DV_D}\int
d^D{\bf r}\exp(-\beta k {r}^2/2) \int_{|h|\leq\delta}d^D{\bf h} \min(1,\exp(-
\beta k(|{\bf r} +{\bf h}|^2-{r}^2)/2))
\end{equation}
where $V_D=\pi^{D/2}/\Gamma(D/2+1)$ is volume of the sphere of unit radius in $D$
dimensions and
\begin{eqnarray}
c_D&=\left(DV_D\int_0^\infty dr r^{D-1}e^{-\beta kr^2/2}\right)^{-1}=
\left(\frac{\beta k}{2\pi}\right)^{D/2}.
\end{eqnarray}

In odd dimensions, the  derivative of Eq.~(\ref{eq:9}) with respect to
$\delta$  can be   written explicitly    by  using  generalized spherical
coordinates
\begin{eqnarray}\label{eq:10}
\frac{d(\delta^D P_{acc}(\delta))}{d\delta} &=\frac{c_D}{ V_D} \int d^D{\bf r}\exp(-\beta k {r}^2/2) \nonumber\\
&\delta^{D-1} \int d\Omega \min(1,\exp(- \beta k(r\delta\cos\phi_1+\delta^2/2))),
\end{eqnarray}
where  $d\Omega  =(\prod_{j=1}^{D-2}(\sin(\phi_j))^{D-1-j}d\phi_j)d\phi_{D-1}$ such that
$\int d\Omega=DV_D$.  The first $D-2$  variables $\phi_j$ are integrated from $0$
to $\pi$, whereas  $\phi_{D-1}$  is integrated from   $0$ to $2\pi$.   If  we
denote  $u=\cos(\phi_1)$,   perform  integration  over  $\phi_2\ldots\phi_{D-1}$,
and introduce the variable $v=r\sqrt{\frac{\beta k}{2}} $
Eq. (\ref{eq:10}) can be rewritten as
\begin{eqnarray}\label{eq:11}
\fl\frac{d(\xi^D P_{acc}(\xi))}{d\xi }&= D\xi^{D-1}  \frac{2}{\Gamma\left(\frac{D}{2}\right)\int_{-1}^1du (1-u^2)^{(D-3)/2}}\times  \nonumber\\
\lo \times & \int_0^{+\infty }  v^{D-1}
dv e^{-v^2} 
\int_{-1}^1du (1-u^2)^{(D-3)/2} \min(1,\exp(- (2\xi vu+ \xi^2)))
\end{eqnarray}
where $\min(1,\exp(-(2\xi  vu+ \xi^2)))=\exp((-(2\xi  vu+ \xi^2))$
for $v<\xi  /2$ with $-1<u<1$ and for $v>\xi  /2$ with  $ -\xi /(2v)<u<1$ and $
\min(1,\exp(- (2\xi    vu+ \xi^2)))=1$ for  $r>\xi   /2$    with  $-1<u<
-\xi /(2v)$. Using that  $\int_{-1}^1du (1-u^2)^{(D-3)/2}=\frac{\Gamma \left(\frac{D-1}{2}\right)}{\Gamma \left(\frac{D}{2}
\right)}\sqrt \pi$, Eq.~(\ref{eq:11}) then becomes
\begin{eqnarray}\label{eq:12}
\fl\frac{d(\xi^D P_{acc}(\xi))}{d\xi }=&\frac{2D\xi^{D-1}}{\sqrt{\pi }\Gamma\left(\frac{(D-1)}{2}\right)}\left( \int_0^{\xi  /2} dv v^{D-1}
 e^{- v^2}  \int_{-1}^1du (1-u^2)^{(D-3)/2} \exp(- ( \xi^2+2\xi 
vu))\right.\nonumber\\
\lo+&\int_{\xi  /2}^{+\infty}  v^{D-1} dv e^{- v^2}\left[\int_{-1}^{-\xi /(2v)}du
(1-u^2)^{(D-3)/2} \right.+\nonumber\\
\lo +&\left.\left.\int_{-\xi /(2v)}^1du (1-u^2)^{1/2}\exp((-( \xi^2+2\xi  vu)))
\right]\right).
\end{eqnarray}

After some calculation (see Appendix A) one obtains, 
\begin{eqnarray}\label{eq:13}
\frac{d(\xi^D P_{acc}(\xi))}{d\xi }&=D\xi^{D-1}\left(1-erf\left(\frac{\xi}{2}\right)\right)
\end{eqnarray}
which gives, for instance, in three dimensions
\begin{equation}\label{eq:37}
P_{acc}(\xi)=      1-erf\left(\frac{\xi}{2}\right)         +         \frac
{8}{\sqrt{\pi}\xi^3}\left(1-\left(1+\frac{\xi^2}{4}\right)
\exp\left(-\frac{\xi^2}{4}\right)\right)
\end{equation}
Figure \ref{fig:1} shows the  acceptance ratio  $P_{acc}$ versus $\delta$ in one, two
and  three   dimensions.

A similar calculation for generating function $Z_D(\lambda,\xi)$  leads to
\begin{equation}\label{eq:36}
\frac{\partial(\xi^D Z_D(\tilde{\lambda},\xi))}{\partial\xi }=D\xi^{D-1}e^{-\tilde{\lambda}\xi}\left(1-erf\left(\frac{\xi}{2}\right)\right),
\end{equation}
and to
\begin{equation}\label{eq:35}
Z_D(\tilde{\lambda},\xi)=D(-\xi)^{1-D}\frac{\partial^{D-1}}{\partial\tilde{\lambda}^{D-1}}Z_{D=1}(\tilde{\lambda},\xi),
\end{equation} 
where the expression for $Z_{D=1}(\tilde{\lambda},\xi)$ is given in
Eq.~(\ref{eq:8}). Since $P_{acc}(\xi)=Z_{D}(\tilde{\lambda}=0,\xi)$, it follows
from Eq.~(\ref{eq:35}) that the acceptance ratio in $D$ dimensions is
equal, up a factor $D(-\xi)^{1-D}\sqrt{\frac{\beta k}{2}}$, to the mean
squared displacement $<(\Delta x)^2>$ in one dimension: compare
Eq.~(\ref{eq:37}) to Eq.~(\ref{eq:32}).

Although straightforward, the algebra rapidly becomes tedious, and we
only illustrate the results by giving the expression of the mean
squared displacement   $<(\Delta r)^2>$ in three dimensions
\begin{eqnarray}
\fl\frac{\beta k}{2}\ <(\Delta r)^2>&=\frac{\partial^2}{\partial\tilde{\lambda}^2}Z_{D=3}(\tilde{\lambda},\xi)|_{\tilde{\lambda}=0}
\nonumber\\&= \frac{3}{\xi^2 }\left(\frac{\beta k}{2}\right)^2<(\Delta x)^4>_{D=1}\nonumber\\
 &=\frac{12}{5
}\left[\frac{\xi^2}{4}\left(1-erf\left(\frac{\xi}{2}\right)\right) 
+\frac{16}{\sqrt{\pi}\xi^2 }\left(1-\left(1+\frac{\xi^2}{4}
+\frac{\xi^4}{32} \right)e^{\left(-\frac{\xi^2}{4}\right)}\right)\right].
\end{eqnarray}
The mean squared and mean absolute displacement are plotted versus the
acceptance  ratio  $P_{acc}$  in one,  two   and  three  dimensions in
Figs.~\ref{fig:2} and ~\ref{fig:3}.  Note that  the maximum is shifted
to  the left, i.e.,  to the smallest values   of the acceptance ratio,
when the space dimension increases.
\subsection{Uniform radius sampling}
The  acceptance ratio  $P_{acc,w}$  in  $D$ dimensions  can  be
expressed as
\begin{eqnarray}\label{eq:15}
P_{acc,w}(\delta)&=c_D\int_D   d^D{\bf  r}\exp(-\beta  k   {\bf r}^2/2)  \nonumber\\
&\int_{|h|\leq\delta}d^D{\bf     h}  P_w(h)\min(1,\exp(- \beta    k(|{\bf   r}  +{\bf
h}|^2-{r}^2)/2))
\end{eqnarray}
where  $P_w(h)$  is  the weighted  probability.   For a  uniform
distribution in radius, $h^{D-1}P_w(h)=(DV_D\delta)^{-1}$. Using the method
developed in the above section, it is straightforward to obtain that
\begin{equation}\label{eq:16}
\frac{d(\xi P_{acc,w}(\delta))}{d\xi }=\left(1-erf\left(\frac{\xi}{2}\right)\right)
\end{equation}
which  shows  that  the  acceptance ratio   is the  same  whatever the
dimension  and explains  the  data collapse observed in  \cite{BKS92}.
Similarly, the generating function $Z(\lambda,\xi)$ can be shown to obey to
the differential equation
\begin{equation}
\frac{\partial Z(\lambda,\xi)}{\partial\xi }=\exp(-\sqrt{\frac{2}{\beta k}}\lambda \xi) \left(1-erf\left(\frac
\xi 2\right )\right), 
\end{equation}
independently of the dimension $D$. This proves that $Z(\lambda,\xi)$ and all
moments such as $<|\Delta r|>$ and  $<(\Delta r)^2>$ are independent of
dimension, as numerically found by BKS\cite{BKS92}.

\section{Dynamic behavior}
In  addition to the exact  results for the static properties presented
above, we  have investigated the dynamic  behavior of the SHO by using
numerical and analytical  approaches. For simplicity, we discuss  only
the unidimensional  case, but the  approach can be generalized  to $D$
dimensions.

The master equation describing the dynamical evolution of the system
during the Monte Carlo simulation, Eq.~(\ref{eq:1}), can be formally
written 
\begin{equation}
\frac{d{\bf P}(t)}{dt}=-L{\bf P}(t)
\end{equation}
where $L$ is   a linear operator acting on  $P$ and the Metropolis
rule, Eq~(\ref{eq:4}), is used for the transition rate. We consider the
spectrum  of  eigenvalues   $\lambda$  of   $L$.  Denoting   $P_\lambda(x)$    the
eigenfunction  associated  with  $\lambda$ and  introducing    $f_\lambda (x)$ via
$P_\lambda(x)=P_{eq}(x)f_\lambda   (x)$,   where   $P_{eq}(x)$    is    given   in
Eq.~(\ref{eq:3}), one can express the eigenvalue equation
\begin{equation}
\lambda P_{eq}(x)f_\lambda (x)=L( P_{eq}(x)f_\lambda (x))
\end{equation}
as 
\begin{equation}\label{eq:17}
\lambda f_\lambda (x)=\frac{1}{2\delta}\int_{-\delta}^{\delta}dh Min(1,e^{-\beta (V(x+h)-V(x))})(f_\lambda (x)-f_\lambda (x+h)).
\end{equation}

Multiplying both  sides  by $\exp(-\beta  V(x))f^*_\lambda(x)$,  where  the star
denotes a complex conjugate, and integrating over $x$ then gives

\begin{equation}\label{eq:18}
\lambda =\frac{1}{4\delta}\frac{\int_{-\infty}^{+\infty}dx\int_{-\delta}^{\delta }dh
Min(e^{-\beta V(x)},e^{-\beta V(x+h)})|f_\lambda (x+h)-f_\lambda(x)|^2}{
\int_{-\infty}^{+\infty}dx e^{-\beta V(x)}|f_\lambda (x)|^2 }.
\end{equation}

As anticipated  for a Markov  process satisfying detailed balance, one
deduces from the  above formula  that  all  eigenvalues are real   and
positive; the smallest eigenvalue is $\lambda_0=0$ and it is associated with
$f_0(x)=constant\neq     0$.     One  need       consider only   real
eigenfunctions. Moreover, the eigenvalues  can be sorted according
to the symmetry of the associated eigenfunctions:  it is easy to check
that the eigenfunctions are  either even or  odd functions of $x$, due
to the fact that the potential $V(x)$ is an even function of $x$.

Any solution of the master equation can be expanded as
\begin{equation}\label{eq:19}
P(x,t)=P_{eq}(x)(1+\sum_{\lambda  >0 }c_\lambda  f_\lambda (x)e^{-\lambda  t}),
\end{equation}
and   a  similar expansion applies    to  the conditional  probability
$P(x,t|x_0,0)$    from which  one     can  compute any  time-dependent
correlation function.   The long-time  kinetics governing the approach
to  equilibrium in $P(x,t)$  and   in  any  correlation function    is
characterized by the  smallest non-zero    eigenvalue for which    the
amplitude,   i.e., the projection   of   $P(x,0)$, or of  the  dynamic
observable, onto the relevant eigenfunction, does not vanish.

Since,  according to Eq.~(\ref{eq:18}),  the eigenvalues are expressed
as the   ratio of two  positive   quadratic functionals (that   in the
denominator being also  definite),   one  can use  the   Rayleigh-Ritz
procedure  to     find   a    variational     upper-bound  for     the
eigenvalues\cite{D62,A85}. Consider first the smallest non-zero
eigenvalue $\lambda_1$. For any real function  $\phi(x)$ which is both
normalized and orthogonal to $f_0(x)$, i.e., satisfies
\begin{eqnarray}
\int_{-\infty}^\infty P_{eq}(x)\phi (x)^2dx&=1\\
\int_{-\infty}^\infty P_{eq}(x)\phi (x)dx&=0,
\end{eqnarray}
one has the inequality
\begin{equation}
\fl \lambda_1\leq \lambda_1[\phi]=\frac{1}{4\delta}\int_{-\infty}^{+\infty}dx\int_{-\delta}^{\delta }dh
Min(e^{-\beta V(x)},e^{-\beta V(x+h)})|\phi  (x+h)-\phi (x)|^2.
\end{equation}

A convenient   choice  of  trial  functions  is  provided  by   linear
combinations   of Hermite polynomials   $H_n(\xi )$   (where, as in  the
previous sections $\xi=\sqrt{\frac{\beta k}{2}}\delta$), with $n> 0$, since these
form, up to  a  trivial multiplicative  factors, an  orthonormal basis
with    respect   to  the    weight  function   $\exp(-\beta   V(x))$ with
$V(x)=(1/2)kx^2$. For  $\lambda_1$,    which is  associated with    an   odd
eigenfunction, one need consider only the odd polynomials $H_{2n+1}(\xi
)$, $n\geq 0$.

The  simplest  estimate  of  $\lambda_1$     is   provided by      taking
\begin{equation}
\phi(\xi )=\frac{H_1(\xi)}{\sqrt{2\sqrt{\pi}}}=\sqrt{\frac{2}{\sqrt{\pi}}}\xi ,
\end{equation}
 which gives
\begin{equation}\label{eq:20}
 \lambda_1[\phi]=\frac{4}{3}
\left(\frac{\xi^2}{4}\left(1-erf\left(\frac{\xi}2\right)\right)\right)
+\frac{2}{\sqrt{\pi}\xi}(1-\left(1+\frac{\xi^2}{4}\exp\left(\frac{-\xi^2} 4\right)\right)
\end{equation}
With this
choice of $\phi(\xi )$, $ \lambda_1[\phi]$ simply reduces to the mean squared
displacement $<(\Delta x)^2>$ multiplied by $\left(\frac{\beta k}{2}\right)$
(see Eq.~(\ref{eq:8})). One then derives from section 2 that $
\lambda_1[\phi]$ versus $\xi$ passes though a maximum for $\xi\simeq 2.611648$, which
corresponds to an acceptance ratio of $P_{acc}=0.41767$.

A better estimate of $\lambda_1$ can be obtained by using a linear
combination of $H_1(\xi )$ and $H_3(\xi )$: 
\begin{equation}
\phi(\xi ;\theta)=\left[\frac{\cos(\theta)}{\sqrt{2\sqrt{\pi}}}H_1(\xi )+
\frac{\sin(\theta)}{\sqrt{48\sqrt{\pi}}}H_3(\xi )\right]
\end{equation}
where only one independent parameter $\theta$ appears due to the
normalization condition. The best bound is determined by minimizing
the expression $ \lambda_1[\phi]$ with respect to $\theta$:
\begin{equation}\label{eq:21}
\frac{\partial\lambda_1[\phi]}{\partial\theta}=0.
\end{equation}

\begin{figure}
\begin{center}
\resizebox{10cm}{!}{\includegraphics{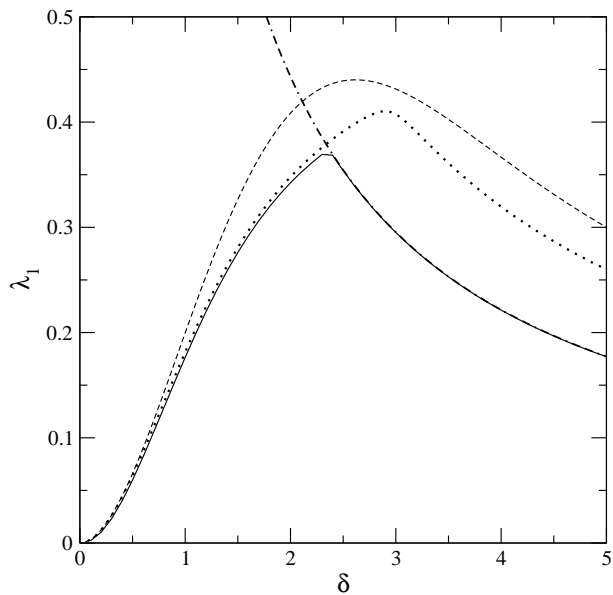}}

\caption{$\lambda_1$ versus $\xi $ . The full curve was obtained by
numerical diagonalization  of the master   equation. The dashed  curve
corresponds to   the zeroth-order  estimate,  Eq.~(\ref{eq:20}),   the
dotted curve   corresponds to the  solution  of the  first-order trial
function, Eq.~(\ref{eq:21})), and the dash-dot curve corresponds to the
exact asymptotic behavior, Eq.~(\ref{eq:31}). }\label{fig:4}
\end{center}
\end{figure}

\begin{figure}
\begin{center}
\resizebox{10cm}{!}{\includegraphics{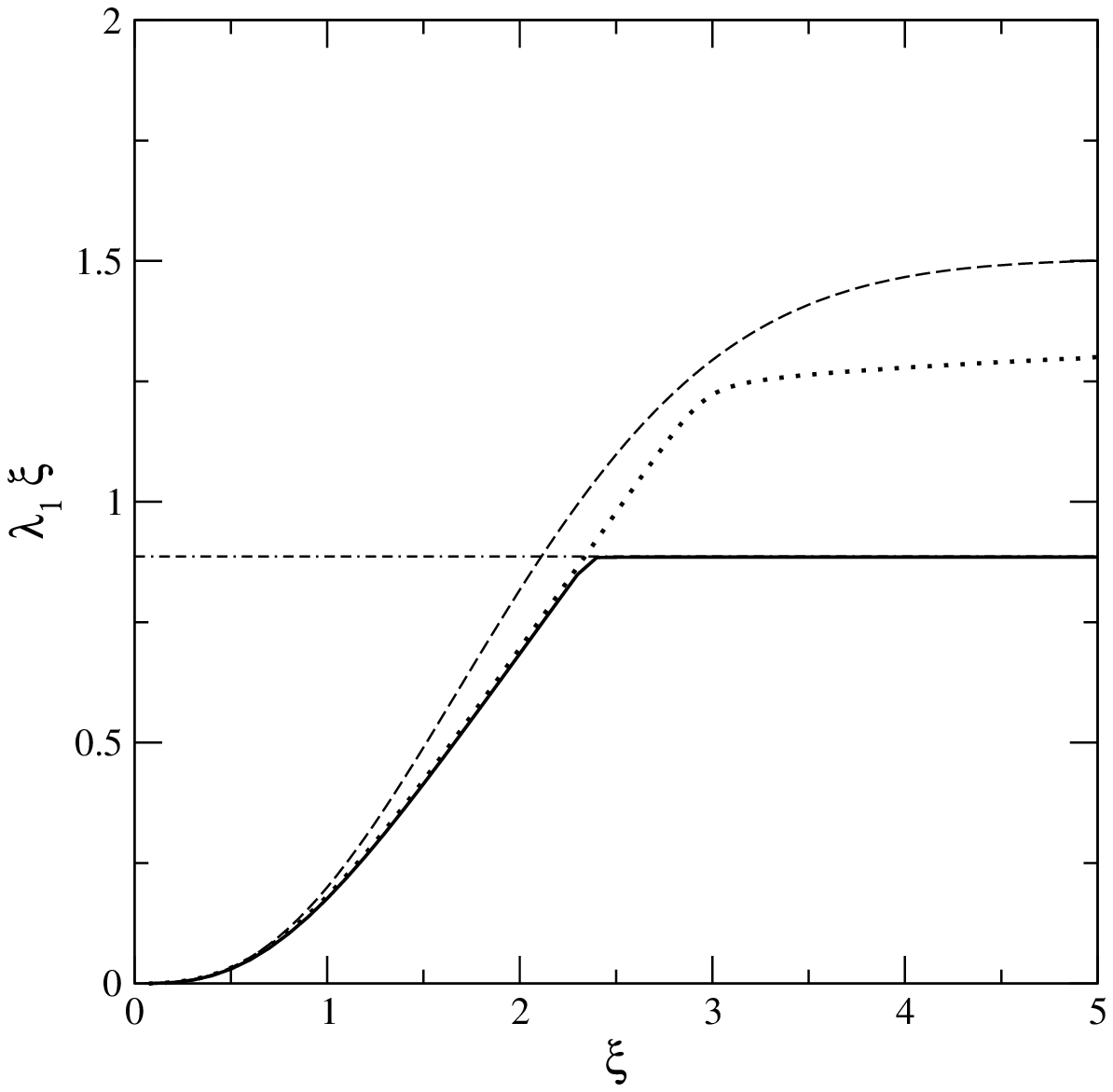}}

\caption{Same as Figure \ref{fig:4} except $\xi \lambda$ versus $\xi$.}\label{fig:5}
\end{center}
\end{figure}

The   result  is a  lengthly   algebraic  formula that   is plotted in
Figs.~\ref{fig:4} and \ref{fig:5}, together   with the expression   in
Eq.~(\ref{eq:20}).

An improved estimate of $\lambda_1$ can be derived by noting that at
large $\xi $, $\lambda_1$ is inversely proportional to $\xi $. Actually, one can
show that this is true for all eigenvalues except $\lambda_0=0$. By
considering Eq.~(\ref{eq:17}) in the limit where $x$ goes to zero, one
arrives at the result (see Appendix B)
\begin{equation}\label{eq:31}
\lambda(\xi)\sim \frac{\sqrt{\pi}}{2\xi }+ 0(e^{-\xi^2}),
\end{equation}
valid for large $\xi $. Note that the correction terms are very small as
soon as $\xi\geq 3$. One can build an estimate of $\lambda_1(\xi)$ by using the
piecewise function that is equal to $ \frac{\sqrt{\pi}}{2\xi }$ for $\xi\geq
\xi^*$ and is equal to $ \lambda_1[\phi]$ obtained for a linear combination of
$H_1$ and $H_3$ (see above) for  $\xi \leq \xi^*$, where $ \xi^*$ is the value
at which  $\lambda_1[\phi]= \frac{\sqrt{\pi}}{2\xi }$. $\lambda(\xi)$ is then maximum for
$\xi=\xi^*\simeq 2.35$; the corresponding value of the acceptance ratio is
$P_{acc}\simeq 0.56$. The estimate is shown in Figs.~\ref{fig:4} and
\ref{fig:5}.

\begin{figure}
\begin{center}
\resizebox{10cm}{!}{\includegraphics{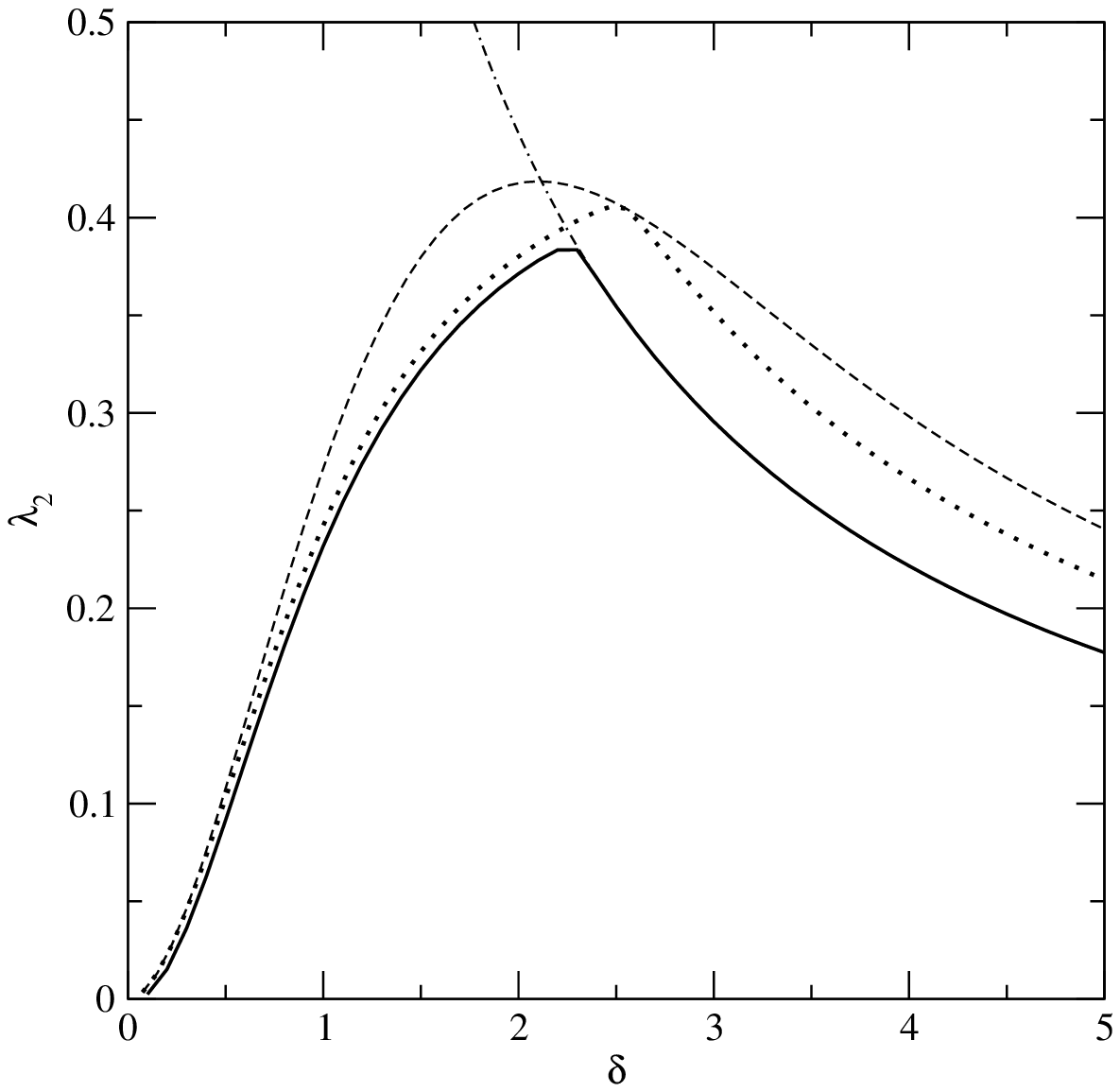}}

\caption{$\lambda_2$ versus $\xi $ . The full curve was obtained by
numerical diagonalization  of  the master equation.   The dashed curve
corresponds to the zeroth order estimate, ($\theta=0$ in Eq.~(\ref{eq:22})),
the  dotted curve  correspond to the   solution  of the  first order trial
function, Eq.~(\ref{eq:22}) and the dash-dot curve corresponds to the
exact asymptotic behavior, Eq.~(\ref{eq:31}).  }\label{fig:6}
\end{center}
\end{figure}

\begin{figure}
\begin{center}
\resizebox{10cm}{!}{\includegraphics{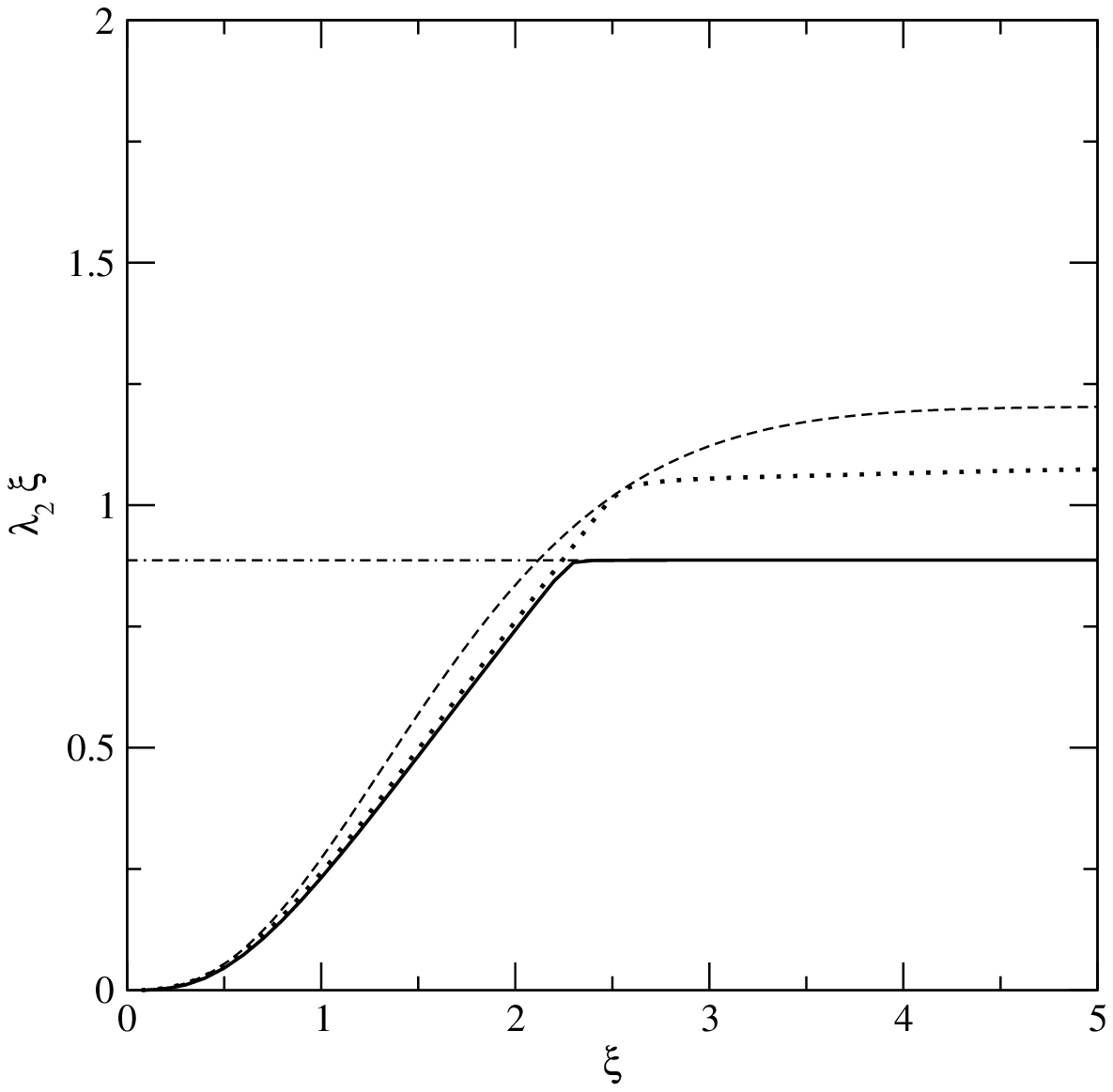}}

\caption{Same as Figure \ref{fig:6} except $\xi \lambda_2$ versus $\xi$}\label{fig:7}
\end{center}
\end{figure}

In order to compare our results  to the BKS  paper \cite{BKS92}, it is
necessary to calculate the second eigenvalue $\lambda_2$. The trial function
is then chosen in a subspace  orthogonal not only to $f_0(\xi )=constant$
but also  to the  eigenfunction associated  with $\lambda_1$.  A  convenient
choice is provided   by (normalized) linear  combinations   of the even
Hermite polynomials $H_{2n}(\xi )$ with  $n\geq 1$.   An
estimate of $\lambda_2$  can be obtained by using   a linear combination  of
$H_2(\xi )$ and $H_4(\xi )$:
\begin{equation}\label{eq:22}
\phi(\xi ;\theta)=\left[\frac{\cos(\theta)}{2\sqrt{2\sqrt{\pi}}}H_2(\xi )+
\frac{\sin(\theta)}{8\sqrt{6\sqrt{\pi}}}H_4(\xi )\right]
\end{equation}
and the result is     shown in Figs.~\ref{fig:6}   and   ~\ref{fig:7},
together      with  the zeroth-order   approximation   obtained   with   only
$H_2(\xi)$ and the improved estimate taking into
account the large-$\xi$ behavior.

In addition to the above analytical estimates, we have also performed
a numerical study of the spectrum of eigenvalues of the master
equation, Eq.~(\ref{eq:1}). The latter has been discretized in
$x$-space by taking a constant step size $\Delta$, which leads to a matrix
form,

\begin{eqnarray}
\frac{d{\bf P}(t)}{dt}=-{\bf W}{\bf P}(t)
\end{eqnarray}
where ${\bf P}$ is a vector with components $ P_i(t)=P(x_i,t)$ and the
elements $W_{ij}$ of the matrix ${\bf W}$ are such that
\begin{eqnarray}
W_{ii}&=  \Delta \sum_{j=i-Nh,j\neq i}^{i+Nh}W(i\to j)\nonumber
\\W_{ij}&= -\Delta W(j\to i)
\end{eqnarray}
and $W(i\to  j)=Min(1,e^{-\beta(V(x_j)-V(x_i))})$.  The eigenvalues  $\lambda$ can
then be obtained via an exact  numerical diagonalization of the matrix
$W$. In practice, convergence is obtained for a unidimensional lattice
of $400$ sites,  where the $x$-range  is $[-10,10]$, and $\xi $ goes from
$0$ to $5$. One checks that the lowest eigenvalue is equal to zero and
corresponds to the  equilibrium state, and  that all other eigenvalues
are real and strictly positive and behave as $\sqrt{\pi}/(2\xi)$ for large
enough $\xi$.  The results for  the first  non-zero eigenvalue $\lambda_1$ and
$\lambda_2$  are   displayed   in    Figs.~\ref{fig:4}   and   ~\ref{fig:6},
respectively. One can see that the best analytical estimate described
above  (linear combination    of two Hermite   polynomials  plus exact
asymptotic behavior at  large $\xi$), is  in excellent agreement  with the
numerical value in both cases.

\begin{figure}
\begin{center}
\resizebox{10cm}{!}{\includegraphics{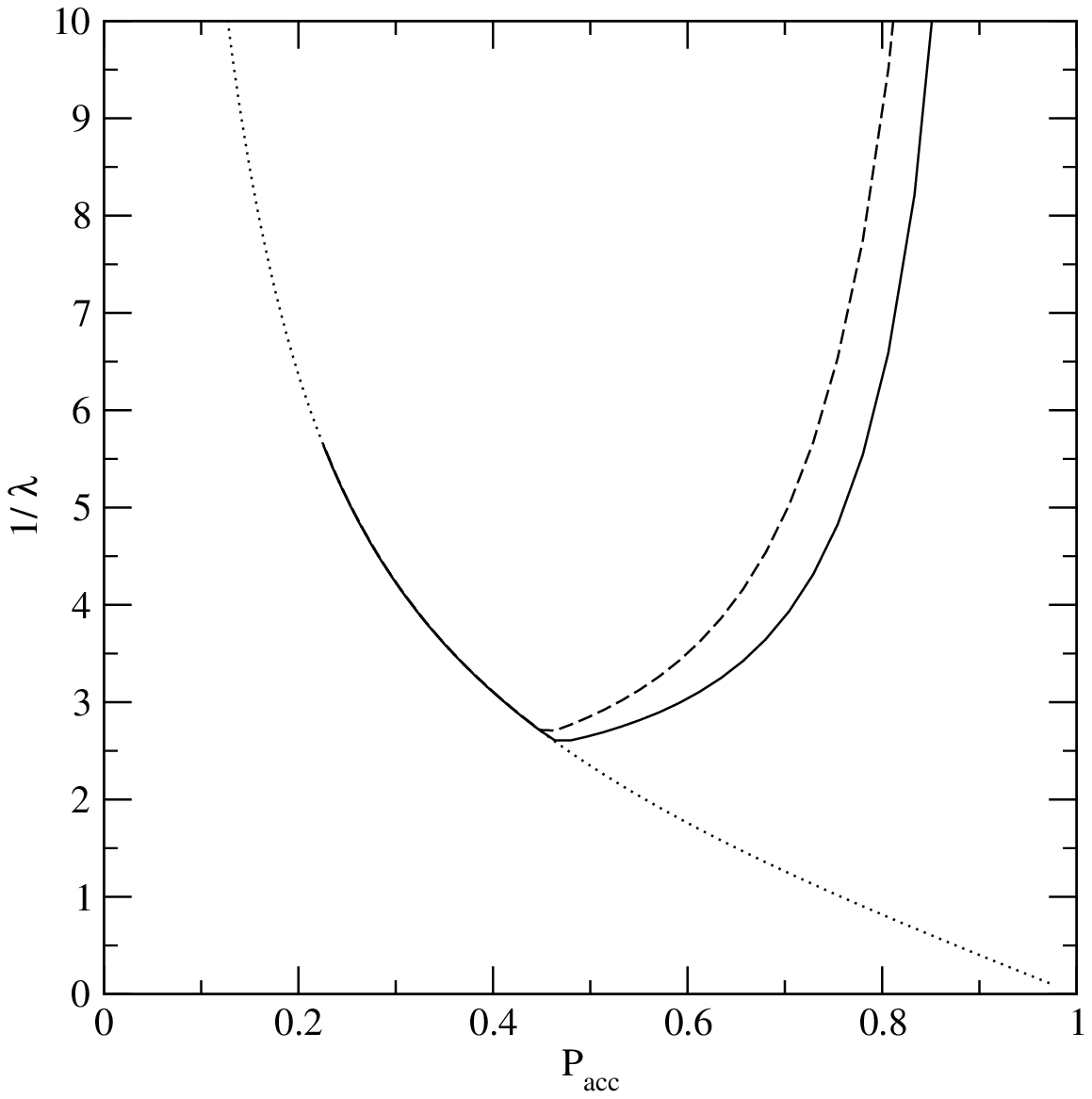}}

\caption{Inverse of the first nonzero eigenvalues, $1/ \lambda_1$ (dotted
curve) and $1/ \lambda_2$ (full curve), versus the acceptance
probability $P_{acc}$}\label{fig:8}
\end{center}
\end{figure}

The above  analysis allows us    to derive by  analytical means  the
simulation result obtained by   BKS\cite{BKS92} for the  dependence of
the characteristic time $\tau$ of the  energy-energy correlation function on
the acceptance ratio: approximating  $\tau$ by $1/ \lambda_2$ (since the energy
$V(x)$  is  an even   function of $x$,   its  projection on  the first
eigenfunction  associated  with $\lambda_1$  vanishes), using  for $\lambda_2$ our
best analytical estimate, and combining this with the exact result for
the acceptance  ratio $P_{acc}$ in section 2  lead  to the full curve
plotted   in Fig.~\ref{fig:8}; the   time   $\tau$  is minimum for    the
acceptance  ratio close  to $0.47$,  as  found by  BKS\cite{BKS92} ($\simeq
0.50$).

In   Fig.~\ref{fig:8}, we  have   also   plotted  $1/\lambda_1$ versus   the
acceptance ratio:  it is minimum  for $P_{acc}=0.45$.  Note that using
the results of this and   the preceding sections, one can   rigorously
show  that  the correlation time  $\tau$, no  matter  how it is precisely
defined,    diverges    as $1/P_{acc}$    when    $P_{acc}\to0$  and  as
$1/(1-P_{acc})$ when $P_{acc}\to1$.

\section{Conclusion}
We  have obtained exact results  concerning  Metropolis Algorithms for
the displacement  of a particle in the  simple harmonic potential. Our
analysis provides a  theoretical explanation of  the numerical results
obtained by BKS\cite{BKS92}.  In particular,  we show that the results
become  independent of the space  dimension when  the successive trial
moves  are sampled according to  a Metropolis algorithm with a uniform
distribution in radius   (instead of  volume). This  rationalizes  the
search for efficient Monte Carlo methods for the simulation of systems
with  intrinsic  inhomogeneity   and anisotropy   such  as  biological
molecules\cite{BKS92,S02}
\appendix
\section{Acceptance probability}
The derivation of Eq.~(\ref{eq:13}) can be done from Eq.~(\ref{eq:12})
after some manipulations whose details are given here. Let us denote
$I_D$ and $J_D$ as
\begin{equation}
I_D= \int_0^{\xi  /2} dv v^{D-1}
 e^{- v^2}  \int_{-1}^1du (1-u^2)^{(D-3)/2} \exp(- ( \xi^2+2\xi 
vu))
\end{equation}
\begin{eqnarray}
 J_D&=\int_{\xi  /2}^{+\infty} dv v^{D-1}  e^{- v^2}\left[\int_{-1}^{-\xi /(2v)}du
(1-u^2)^{(D-3)/2} \right.+\nonumber\\
  &+\left.\int_{-\xi /(2v)}^1du (1-u^2)^{(D-3)/2}\exp((-( \xi^2+2\xi  vu)))
\right].
\end{eqnarray}
Changing the variable $u$ to $y=vu$ and $v$ to $t=v^2-y^2$ leads to
the following relations
\begin{equation}
I_D=\int_0^{\xi/2}dy\frac{\exp(-(\xi+y)^2)+\exp(-(y-\xi)^2)}{2}\int_0^{\xi^2/4-y^2}dt\, t^{(D-3)/2}e^{-t}
\end{equation} 
\begin{eqnarray}
J_D&=\int_{\xi/2}^{+\infty }dy\frac{\exp(-y^2)}{2}\int_{0}^{+\infty}dt\,t^{(D-3)/2}e^{-t}
\nonumber\\
&+\int_0^{\xi/2}dy \frac{\exp(-(y-\xi)^2)}{2}\int_{\xi^2/4-y^2}^{+\infty}dt\,t^{(D-3)/2}e^{-t}
\nonumber\\
&+\int_0^{\xi/2}dy
\frac{\exp(-(y+\xi)^2)}{2}\int_{\xi^2/4-y^2}^{+\infty}dt\,t^{(D-3)/2}e^{-t}
\nonumber\\
&+\int_{\xi/2}^{+\infty }dy
\frac{\exp(-(y+\xi)^2)}{2}\int_{0}^{+\infty}dt\,t^{(D-3)/2}e^{-t}
\end{eqnarray}
Using that $\int_{0}^{+\infty}dt\,t^{(D-3)/2}e^{-t}=\Gamma((D-1)/2)$, one obtains
that
\begin{equation}\label{eq:14}
I_D+J_D=\frac{\sqrt{\pi}}{2}\Gamma\left(\frac{D-1}{2}\right)
\left(1-erf\left(\frac{\xi}{2}\right)\right)
\end{equation}
Inserting Eqs.~(\ref{eq:14})  in Eq.~(\ref{eq:12})
leads to Eq.~(\ref{eq:13}). 

\section{Asymptotic behavior of the eigenvalues for large $\delta$}
Consider Eq.~(\ref{eq:17}) in the limit $x\to0$ ; one
obtains after rearranging the various terms:
\begin{eqnarray}\label{eq:23}
\fl\lambda f_\lambda (x)&=\frac{1}{2\delta}\int_{-\delta}^{\delta}dh \exp(\frac{-\beta k}{2} ((x+h)^2-x^2))
(f_\lambda (x)-f_\lambda (x+h))\nonumber\\
&+\frac{1}{2\delta}\int_{-2x}^{0}dh(1-\exp(\frac{-\beta k}{2} ((x+h)^2-x^2)))(f_\lambda (x)-f_\lambda (x+h)).
\end{eqnarray}
The second term of the right-hand side of Eq.~(\ref{eq:23}) is at most of
order $x^3|f_\lambda (x)|$ and is always negligible so that one can rewrite
Eq.~(\ref{eq:23}) as
\begin{equation}\label{eq:24}
\fl\left(\lambda\delta - \int_{0}^{\delta }dh e^{\frac{-\beta k}{2}h^2}+O(x^2)\right)  f_\lambda (x)\simeq -\frac{e^{\frac{-\beta k}{2}x^2}}{2}\int_{-\delta}^{\delta}dh \exp(\frac{-\beta k}{2} (x+h)^2)f_\lambda (x+h).
\end{equation}
Shifting the  variable from $h$ to $x+h$  in the integral of the r.h.s
of    Eq.~(\ref{eq:24}) and using the    orthogonality of $f_\lambda (x)$ to
$f_0(x)=constant$ for all non-zero eigenvalues $\lambda$, i.e., 
$\int_{-\infty }^{+\infty }dx \exp(\frac{-\beta k}{2} x^2)f_\lambda (x)=0$, leads to the
following expression,
\begin{equation}\label{eq:25}
\fl\left(\lambda\delta - \sqrt{\frac{\pi }{2\beta k}}+ O(x^2)\right)  f_\lambda (x)\simeq \frac{1+
O(x^2)}{2}\left[\int_{x+\delta}^{+\infty }dh e^{\frac{-\beta kh^2}{2}} f_\lambda (h)+\int^{x-\delta}_{-\infty }dh e^{\frac{-\beta kh^2}{2}} f_\lambda (h)\right]
\end{equation}
for any non-zero $\lambda$.
The r.h.s. of Eq.~(\ref{eq:25}) can be Taylor expanded, which gives
\begin{eqnarray}\label{eq:26}
\fl \left(\lambda\delta - \sqrt{\frac{\pi }{2\beta k}}+ O(x^2)\right)  f_\lambda (x)&\simeq \frac{1}{2}
\left[\int_{\delta}^{+\infty }dh e^{\frac{-\beta kh^2}{2}}( f_\lambda (h)+f_\lambda(-h))\right.\nonumber\\
&\left. -x
e^{\frac{-\beta kh^2}{2}}( f_\lambda (h)-f_\lambda(-h))+ 0(x^2)\right].
\end{eqnarray}
If the eigenfunctions $f_\lambda$ is an even function of $x$, one then
derives that $ f_\lambda (x)=f_\lambda(0)+ O(x^2)$ with $f_\lambda(0)\neq 0$ and
\begin{equation}\label{eq:27}
\left(\lambda\delta - \sqrt{\frac{\pi }{2\beta k}}\right)=\int_{\delta}^{+\infty }dh
\exp(\frac{-\beta kh^2}{2}) \frac{ f_\lambda (h)}{ f_\lambda (0)},
\end{equation}
which after introducing $\xi=\sqrt{\frac{\beta k}{2}}\delta$ can be rewritten as
\begin{equation}\label{eq:28}
\lambda= \frac{\sqrt{\pi }}{2\xi }+\sqrt{\frac{2}{\beta k}}\int_{\xi }^{+\infty }dh
\exp(\frac{-\beta kh^2}{2}) \frac{ f_\lambda (\sqrt{\frac{2}{\beta k}}h)}{ f_\lambda (0)}.
\end{equation}
If the eigenfunction is an odd function of $x$, one has that $f_\lambda
(x)=f'_\lambda(0)x(1 + O(x^2)$ with $f'_\lambda(0)\neq 0$ and 
\begin{equation}\label{eq:29}
\lambda= \frac{\sqrt{\pi }}{2\xi }
 \frac{ f_\lambda (\sqrt{\frac{2}{\beta k}}\xi )}{ f_\lambda (0)}\exp(-\xi^2).
\end{equation}
From Eqs.~(\ref{eq:28}) and (\ref{eq:29}), one immediately obtains
that all non-zero eigenvalues behave as 
\begin{equation}
\lambda\sim  \frac{\sqrt{\pi }}{2\xi }
 + O(\exp(-\xi^2))
\end{equation}
when $\xi \to+\infty$, since $f_\lambda(x)$ diverges more slowly than $e^{x^2}$ when
$x\to+\infty$.
\section*{References}


\end{document}